\documentclass[screen]{acmart}

\AtBeginDocument{%
  \providecommand\BibTeX{{%
    \normalfont B\kern-0.5em{\scshape i\kern-0.25em b}\kern-0.8em\TeX}}}

\setcopyright{rightsretained}
\copyrightyear{2021}
\acmYear{2021}
\acmDOI{}

\acmConference[]{Workshop Human Aspects of Misinformation at CHI 2021}{May 08--13, 2020}{Yokohama, Japan}
\acmBooktitle{Workshop Human Aspects of Misinformation at CHI 2021, May 08--13, 2020, Yokohama, Japan}
\acmPrice{15.00}
\acmISBN{}

\begin{document}

\title[Helping People Deal With Disinformation]{Helping People Deal With Disinformation - A Socio-Technical Perspective}

\author{Hendrik Heuer}
\affiliation{%
  \institution{University of Bremen}
  \city{Bremen}
  \country{Germany}}
\email{hheuer@uni-bremen.de}

\renewcommand{\shortauthors}{Heuer}

\maketitle

\section{Motivation}

At the latest since the advent of the Internet, disinformation and conspiracy theories have become ubiquitous. Recent examples like QAnon and Pizzagate prove that false information can lead to real violence. Sadly, this link between misinformation and violence has a long history. Anti-semitic conspiracy theories played a central role in the Shoa. In the context of the COVID-19 pandemic, the World Health Organization (WHO) warned that ``misinformation costs lives'' \cite{world_health_organization_managing_2020}. The WHO argued that mitigating the harm from misinformation is necessary to manage the \textit{COVID-19 infodemic}. The recent increase in misinformation in a so-called \textit{Post-Truth Era} can be linked to societal mega-trends such as a decline in social capital, growing economic inequality, increased polarization, declining trust in science, and an increasingly fractionated media landscape \cite{doi:10.1177/1529100612451018}.

My work is focused on disinformation. Following \citet{claire_wardle_age_2014}, I operationalize disinformation as fabricated or deliberately manipulated content like conspiracy theories or rumors. As such, disinformation is a special case of misinformation. While disinformation is connected to an intent to harm, misinformation also includes unintentional mistakes. I believe that disinformation is a challenging, multifaceted phenomenon that requires an appropriate socio-technical response. In my work, I research 1. why people believe in disinformation, 2. how people can be best supported in recognizing disinformation, and 3. what the potentials and risks of different tools designed to fight disinformation are.

My work on disinformation is informed by my background in human-computer interaction and machine learning. The workshop would be a great opportunity for me to discuss the ethical implications of the disinformation detection solutions that I am developing. I would love to discuss ways of enabling fast and effective disinformation detection while ensuring that freedom of speech is protected.

\section{Machine Learning-based Curation Systems}

To understand disinformation in the contemporary media climate, one has to understand social media and the machine learning-based curation systems used on social media. My doctoral thesis provides a socio-technical perspective on users and machine learning-based curation systems \cite{elib_4444}. The thesis presents actionable insights on how ML-based curation systems can and should be explained and audited. Motivated by the role that ML-based curation systems play in the dissemination of disinformation, I examined the user beliefs around such systems in detail. In a recent CSCW paper, I, together with my collaborators, examined how users without a technical background, who regularly interact with YouTube's ML-based curation systems, think the system works \cite{10.1145/3415192}. Our semi-structured interviews with participants from Belgium, Costa Rica, and Germany show that users are aware of the existence of the recommendation system on YouTube, but that users' understanding of the system is limited. This has important consequences for the dissemination of disinformation. With my upcoming work on disinformation, I extend on previous work on how users quantify trust in news on social media \cite{10.1145/3240167.3240172}. In this paper, I identify factors that influence this trust and show that while the majority of users can provide nuanced ratings that correspond to ratings of media experts, a small number of extreme users tend to over- and under-trust in news. A follow-up work focused on the output of ML-based curation systems showed that users can provide trust ratings that distinguish trustworthy recommendations of quality news stories from untrustworthy recommendations \cite{heuer2020fake}. However, a single untrustworthy news story combined with four trustworthy news stories is rated similarly as five trustworthy news stories. This could be the first indication that untrustworthy news stories benefit from appearing in a trustworthy context.

\section{Why People Believe in Disinformation}

To understand why people are vulnerable to disinformation and what people really need to recognize disinformation, I interviewed domain experts from different fields like media science, law, psychology, sociology, political science, and others. I asked the experts why they think people are susceptible to disinformation. My preliminary results show that a variety of influence factors is recognized by the different domain experts. Some experts say that people believe in disinformation because it aligns with their beliefs and because it speaks to their cognitive biases. Other experts think users may lack the education to recognize disinformation. Some users may be tempted to follow recommendations from their peers. Others may simply be overwhelmed by an increasingly complex world and a large amount of information. Certain people may even use disinformation intentionally for personal gain.

\section{How People Can Be Best Supported}

Together with the domain experts, I developed a variety of solutions to the problems they described, both technical and non-technical. These solutions range from formal education and laws to the flagging of content or sources. The solutions include tools that automatically fact-check articles as well as chatbots that train people to recognize disinformation. In an upcoming paper, I map out the design space for solutions that help people recognize disinformation.

Based on the solutions proposed by the experts, I am currently developing a number of other tools, e.g. to support users in assessing the source of a news article, to identify the author of a news article, and to support users in fact-checking the content of an article. I will compare these tools to written checklists and to a setting where the user has no support.

One proposed solution that I investigated in depth was a machine learning-based, automated flagging system that recognized articles based on style. Considering the nature of news as information about recent events, systems based on lexical features are not able to account for new concepts and the changing meaning of words. To mitigate these limitations of an ML-based approach focused on content, I developed a system that can detect disinformation based on the style of a news article, extending on previous lexical approaches \cite{perez-rosas-etal-2018-automatic,wang-2017-liar}. Developing a machine learning system that detects disinformation based on style, rather than content was motivated by previous research on the linguistic and stylistic signals related to misinformation and fact-checking \cite{10.1145/3274351,schuster2020limitations,Khalid_Srinivasan_2020}. The style-based system is able to detect disinformation based on stylistic features with F1-scores of 80 or higher. However, a pilot study in October 2020 provided evidence that explanations like ``The average number of words per sentence is low'', the ``Usage of perceptual words related to hear is low'' and ``The amount of words related to people is high'' are not perceived as helpful by users. Even educated participants had trouble understanding explanations of ML-based systems, a problem that I have previously reported on in the context of ML-based curation systems~\cite{elib_4444} and object recognition systems~\cite{10.1145/3340631.3394873}.

\subsection{Source-Based Disinformation Detection}

My preliminary results indicate that providing information on the reliability of the source of a news story is the most promising direction. The advantage of this approach is that is scalable (the number of new news sources appearing is limited), that it can be explained with little effort (based on the history of the news source), and that it can be integrated into the existing workflow of users. A disadvantage of the approach is that news sources frequently mix correct reporting and false reporting. Therefore, the source assessment as an \textit{all or nothing} approach needs to be situated well. In addition to that, curating the list of reliable and unreliable news sources is a concentration of power and as such politically controversial. To ensure user acceptance, reliable and transparent governance models need to be established, akin to what Wikipedia achieved. I am currently developing a browser extension that augments the interface of existing websites like Facebook and Twitter, which I am planning to evaluate empirically in user studies.

\section{Conclusion}

I want to understand why people believe in disinformation. Based on my theoretical insights, I want to design and develop solutions that support people in recognizing disinformation. My goal is to provide a taxonomy of influence factors that make people prone to believe in disinformation, thus providing a theoretical foundation for the fight against disinformation. I would be happy to discuss the potentials and risks of the different solutions.

\bibliographystyle{ACM-Reference-Format}
\bibliography{references}

\end{document}